\documentclass[epj]{svjour}

\usepackage{graphics}
\usepackage{graphicx}
\usepackage{amssymb}

\usepackage{color}

\begin{document}

\title{Critical behavior and out-of-equilibrium dynamics of a two-dimensional 
Ising model with dynamic couplings}
\author{O. A. Pinto\inst{1} \and F. Rom\'a\inst{2} \and S. Bustingorry\inst{3}}

\institute{
$^1$ Centro de Investigaciones y Transferencia de Santiago del Estero (CITSE),
Universidad Nacional de Santiago de Estero, CONICET
Ruta Nacional 9, Km 1125, Villa el Zanj\'on, 4206, Santiago del Estero, Argentina \\
$^2$ Departamento de F\'{\i}sica, Universidad Nacional de San Luis, INFAP, CONICET, Chacabuco 917, 
D5700BWS San Luis, Argentina \\
$^3$ CONICET, Centro At\'omico Bariloche, R8402AGP San Carlos de Bariloche, R\'{\i}o Negro, Argentina}

\authorrunning{O. A. Pinto \textit{et al.}}
\titlerunning{Critical behavior and out-of-equilibrium dynamics of an Ising model}

\date{Received: date / Revised version: \today }

\abstract{
We study the critical behavior and the out-of-equilibrium dynamics of a two-dimensional 
Ising model with non-static interactions. In our model, bonds are dynamically changing 
according to a majority rule depending on the set of closest neighbors of each spin pair, 
which prevents the system from ordering in a full ferromagnetic or antiferromagnetic state.  
Using a parallel-tempering Monte Carlo algorithm, we find that the model undergoes a 
continuous phase transition at finite temperature, which belongs to the Ising universality class.  
The properties of the bond structure and the ground-state entropy are also studied. Finally, 
we analyze the out-of-equilibrium dynamics which displays typical glassy characteristics 
at a temperature well below the critical one. 
\PACS{
      {75.10.Nr}{Spin-glass and other random models}   \and
      {75.40.Gb}{Dynamic properties (dynamic susceptibility, spin waves, spin diffusion, 
                 dynamic scaling, etc.)} \and
      {75.40.Mg}{Numerical simulation studies}
     } 
} 
\maketitle
\section{Introduction} \label{sec:intro}

Many simple mathematical models showing a complex glassy behavior have been proposed 
in the literature.  From well-known ordered systems, a possible way to obtain its spin glass relative is by introducing 
randomness in the frozen spatial structure of interactions, so as to achieve a highly 
frustrated ground state \cite{Binder1986}. Nevertheless, introducing quenched disorder is not the only possible recipe to obtain some characteristic glassy properties from a given ordered model.
Systems with $p$-body short-range (no random quenched) interaction and $p \ge 3$ 
\cite{Newman1999,Lipowski1997,Swift2000,Buhot2002,Cavagna2003}, 
or incorporating constraints on the maximum permitted number of neighboring particles on 
the lattice \cite{Biroli2001,Krzakala2008}, are valid examples of how it is possible to 
attain a glassy behavior from a model without quenched disorder.  In all these cases, 
in one form or another, Hamiltonians incorporating non-pairwise interactions are invoked 
to accomplish this end.

Instead of starting from a non-disordered model, it is also possible to change 
the glassy properties by modifying a system 
with quenched disorder so that interactions are no longer frozen in space.  
In Ref. \cite{Lazic2001}, for example, based on the two-dimensional (2D) 
Edwards-Anderson $\pm J$ spin-glass model \cite{EA,Toulouse1977}, an Ising system
with mobile bonds was proposed as a viable toy model of vitrification.  By allowing 
bonds to hop to nearest neighbors at the same Glauber Monte Carlo rate as spin flips,
the authors determined that the system has a similar dynamic behavior as found 
in structural glasses (but does not undergo a phase transition at finite temperature \cite{Hartmann2003}).
In addition, a crossover from a liquid-like to a glassy-like
behavior is found for annealed versions of diluted spin-glass models when different 
constraints are imposed to the bonds structure \cite{Fierro2004}.   

With the aim of having an alternative route to glassy behavior, we consider here a strategy 
similar to the one used in Ref. \cite{Lazic2001}, where the interaction bonds are no longer
frozen. Our starting point is also the Edwards-Anderson $\pm J$ spin-glass model and 
we introduce a 2D short-range toy model which exhibits a rich physical behavior. In this model, 
ferromagnetic and antiferromagnetic bonds between pairs of neighboring spins are dynamically 
established from a specific rule defined through the spins configurations surrounding that pair.  
This rule is designed to avoid both, a full ferromagnetic and a full antiferromagnetic ground state. 
We show in this work that such model system presents both a well-defined finite-temperature 
continuous phase transition and non-trivial out-of-equilibrium properties.

The paper is organized as follows. In Sec. \ref{sec:model}, we present  
the model and the Monte Carlo simulation schemes. Section \ref{sec:result}
is devoted to the study of the equilibrium and the ground state properties,
and also to the out-of-equilibrium dynamics.   Finally, in Sec. \ref{sec:conc}
we discuss the obtained results.

\section{The model and the simulation schemes \label{sec:model}} 

The Hamiltonian of the model is 
\begin{equation}
H = - \sum_{(ij)} J_{ij}(\Omega_{ij}) \sigma_i \sigma_j, \label{ham}
\end{equation}
where the sum runs over all pairs of nearest neighbors of a square lattice of linear dimension $L$,
with periodic boundary conditions and the variables $\sigma_i = \pm 1$ represent 
$N=L^2$ Ising spins.  Unlike the Hamiltonian of the 2D Edwards-Anderson model \cite{EA,Toulouse1977},
here the bonds or couplings $J_{ij}$ dynamically depend, according to a majority rule, 
on the closest neighborhood of the pair $\sigma_i$, $\sigma_j$. 
This neighborhood is defined as the six nearest-neighbors spins surrounding the pair $\sigma_i$, $\sigma_j$ 
and is denoted as $\Omega_{ij}$. The coupling $J_{ij}$ are then chosen with the following rule:     
\begin{equation}
J_{ij}(\Omega_{ij})= \left\{ 
\begin{array}{ll}
+1  & \ \ \textrm{if $|m_{ij}| < \frac{1}{2}$}, \\
-1  & \ \ \textrm{if $|m_{ij}| > \frac{1}{2}$},
\end{array} 
\right.
\label{rule}
\end{equation}
where 
\begin{equation}
m_{ij}=\frac{1}{6} \sum_{\Omega_{ij}} \sigma_k.
\end{equation}
In other words, a coupling is chosen to be ferromagnetic (antiferromagnetic), $J_{ij}=+1$ 
($J_{ij}=-1$), if the magnetic order of their environment is mainly antiferromagnetic (ferromagnetic). 
Thus, the majority rule given by Eq. (\ref{rule}) prevents the formation of a perfect ferromagnetic or antiferromagnetic ground state.   

Equilibrium calculations were made using a Monte Carlo parallel-tempering algorithm \cite{Geyer1991,Hukushima1996}. 
It consists in making an ensemble of $R$ replicas of the system, each of which is at temperature 
$T_k$ ($T_1 \geq T_k \geq T_R$).  The basic idea of the algorithm is to independently simulate 
each replica with a single spin-flip dynamics where updates are attempted with a probability given 
by the Metropolis rule \cite{Metropolis}, and to periodically swap the configurations 
of two randomly chosen replicas. A unit of time or parallel-tempering step (PTS), consists of 
a number of $R \times N$ elementary spin-flip attempts followed by only one swap attempt.

The purpose of these swaps is to try to avoid that replicas at low temperatures get stuck in 
local energy minima. Thus, the highest temperature, $T_1$, is set in the high-temperature phase where 
relaxation time is expected to be very short, while the lowest temperature, $T_R$, is set in 
the low-temperature phase. In order to implement the parallel-tempering algorithm, we have chosen 
equally spaced temperatures, i.e. $T_k-T_{k+1}= \left(T_{1} - T_R \right)/(R-1)$.  
Typically, a run starts from a random initial configuration of the ensemble,
half of PTSs are discarded, which is usually enough to reach equilibrium, and averages are performed over the 
remaining simulation steps. More information regarding this Monte Carlo method can be found in Refs. \cite{HartmannRieger,Earl2005,Katgraber2006}. 

Several quantities were numerically computed in order to characterize the equilibrium and 
critical behavior of the described model. In particular, the mean energy per spin was determined as, 
\begin{equation}
e=\frac{\left< H \right>}{N},  
\label{energy}
\end{equation}
where $\langle \cdots \rangle$ represents a thermal average, i. e., the time average throughout 
a Monte Carlo run at temperature $T$.  Also,
the specific heat $C_H$ was sampled from energy fluctuations,
\begin{equation}
C_H= \frac{1}{N T^2} [\langle H^2 \rangle - \langle H \rangle^2].
\label{heat}
\end{equation}
To discuss the nature of the phase transition, the fourth-order energy 
cumulant was computed as
\begin{equation}
U_H(T) = 1 -\frac{\langle H^4\rangle} {3\langle H^2\rangle^2}.
\label{cumH}
\end{equation}
Since the ground state of the system has nonzero net magnetization (see below), the magnetization
\begin{equation}
M = \sum_{i=1}^N \sigma_i,
\label{mag}
\end{equation}
and the mean normalized magnetization
\begin{equation}
m = \frac{\langle \left | M \right | \rangle}{N}
\label{magNor}
\end{equation}
were defined. The magnetization $m$ will be used as an order parameter and it therefore has related quantities such as the susceptibility $\chi$ \cite{LandauBinderBook} and 
the reduced fourth-order Binder cumulant $U_M$ \cite{BinderBook}, which were calculated as
\begin{equation}
\chi = \frac{1}{N T} [ \langle M^2 \rangle - \langle | M | \rangle^2]
\label{susM}
\end{equation}
and
\begin{equation}
U_M(T) = 1 -\frac{\langle M^4\rangle} {3\langle M^2 \rangle^2},
\label{cumM}
\end{equation}
respectively.

At equilibrium we also study the properties of the bond structure.  With this aim, we define
$f$ as the mean fraction of frustrated plaquettes \cite{Toulouse1977}.  A square plaquette is 
frustrated if and only if, the product of the $J_{ij}$ bonds along all four edges of the 
plaquette is a negative number.  Also, we consider the functions $p_{\mathrm{F}}$ and 
$p_{\mathrm{AF}}$ which we define as, respectively, the mean fraction of ferromagnetic and 
antiferromagnetic bonds.

Error bars of equilibrium
quantities are calculated by using standard methods \cite{LandauBinderBook}. It is important
to notice that global moves in the parallel-tempering algorithm significantly reduce 
the critical slowing down and then we can sample in each PTS (the algorithm
reduce the autocorrelation times dramatically, even close to the critical point).
In addition, error bars for the energy and the Binder cumulants are computed using a bootstrap 
method \cite{numerical}.         

Besides equilirium mesures, to explore the low-temperature behavior, we have run out-of-equilibrium
simulations.  A typical protocol is used which consists on a quench at time $t=0$ from a 
random state ($T \to \infty$) to a low temperature $T$.  From this initial condition 
the system is simulated by a standard Glauber dynamics.  Then, the correlation function 
\begin{equation}
C(t,t_w) = \frac{1}{N} \sum_{i=1}^N \langle \sigma_i(t) \sigma_i(t_w) \rangle_0
\label{corr}
\end{equation}
is defined, which depend on both, the waiting time $t_w$ when the measurement begins and a given time 
$t>t_w$. We also computed its associated integrated response function 
\begin{equation}
\rho(t,t_w) = \frac{T}{N} \sum_{i=1}^N \frac{\delta \langle \sigma_i(t) \rangle_h}{\delta h_i}\bigg|_{h \to 0} , 
\label{resp}
\end{equation}
where $h_i$ is a local external field of magnitude $h$, which is switched on only for times $t > t_w$.
In these equations $\langle \cdots \rangle_0$ and $\langle \cdots \rangle_h$ indicate, 
respectively, averages over different thermal histories (different initial configurations and 
realizations of the thermal noise) of the unperturbed and the perturbed system.   
Instead of performing additional simulations with applied fields of small strength, 
the integrated response function (\ref{resp}) was calculated for infinitesimal perturbations 
using the algorithm proposed in Refs.~\cite{Chatelain2003,Ricci-Tersenghi2003}.  
This technique permits us to determine correlation and integrated response functions in a single 
simulation of the unperturbed system. 

At thermodynamic equilibrium, correlation (\ref{corr}) and integrated response (\ref{resp}) 
functions depend on $\tau=t-t_w$ and are related through the fluctuation-dissipation 
theorem (FDT)    
\begin{equation}
\rho(t-t_w) = 1- C(t-t_w).  \label{FDT}
\end{equation}
For a nonequilibrium process, however, the FDT is not fulfilled.  Nevertheless, it has 
been proposed that a generalized quasi-fluctuation-dissipation theorem (QFDT) \cite{Cugliandolo1994,Crisanti2003} 
of the form 
\begin{equation}
\rho(t-t_w) = X(C) \left[ 1- C(t-t_w) \right], \label{QFDT}
\end{equation} 
where $X(C)$ is the fluctuation-dissipation ratio, should be obeyed by any physical model.

Finally, it is worth to mention that for out-of-equilibrium 
simulations all thermal histories
are totally independent of each other, and then errors bars are simply calculated as the 
standard deviation divided by the square root of the number of runs.  

\section{Numerical results} \label{sec:result}

In this section we present the numerical results.  We start by analyzing the critical behavior
of the system at intermediate temperatures.  Then, the bond structure properties in a wider temperature range and its relation with the ground-state configurations are studied.  Finally, in the lower temperature region where for large lattice sizes equilibrium calculations are not possible, we study the 
out-of-equilibrium dynamics of the model.  
For simplicity, error bars are omitted in the figures since those are smaller than data points.    

\subsection{Equilibrium phase transition}

Simulations parameters were chosen after making a typical equilibration 
test \cite{Katzgraber2006}, by studying how the numerical results vary when the number of 
PTSs are successively increased by factors of 2. We require that the last three results 
for all observables agree within the error bars.
Note that, since the system is not disordered, it should be enough to perform 
a single Monte Carlo run.  Nevertheless, we choose to calculate average values along some paths
generated with different initial states and random numbers, and thus to minimize the statistical errors.  

We have simulated ensembles of $R=200$ replicas, with $T_1=5.0$ and $T_R=1.0$ (temperatures 
are given in units of $1/k_B$, where $k_B$ is the Boltzmann's constant).  
Lattice sizes ranging from $L=20$ to $L=80$ were studied using in all cases $10^6$
PTSs and, to calculate the equilibrium values of different observables, we have also performed 
averages over few independent runs ($10^2$ for $L=20$ and only $10$ for the biggest size $L=80$). 

\begin{figure}[t]
\begin{center}
\includegraphics[width=7cm,clip=true]{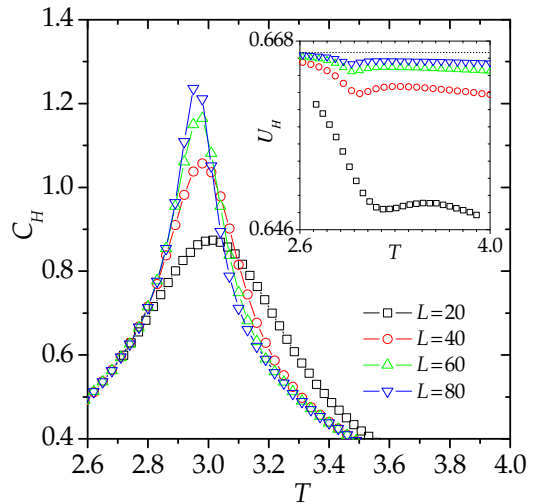}
\caption{Temperature dependence of the heat capacity and the energy cumulant (inset),
for different lattice sizes as indicated.}
\label{figure1} 
\end{center}
\end{figure}

Figure \ref{figure1} shows the heat capacity as function of $T$.  A peak around $T \approx 2.94$,
whose intensity increases with increasing lattice size, indicates the possibility of a phase transition
at that temperature.  As a first step, we analyze the behavior of the energy cumulant.  It is 
well-known that the finite-size analysis of this quantity is a simple and direct way to determine 
the order of a phase transition \cite{Binder1984,Challa1986,Vilmayr1993}.  The curves of $U_H$ versus
$T$ are shown in the inset of Figure \ref{figure1}. It can clearly be observed that the minima 
in the energy cumulants tend to $2/3$ as the lattice size is increased. This behavior is typical of
a continuous phase transition, because it indicates that the latent heat is zero in the thermodynamic 
limit. 

\begin{figure}[t]
\begin{center}
\includegraphics[width=7cm,clip=true]{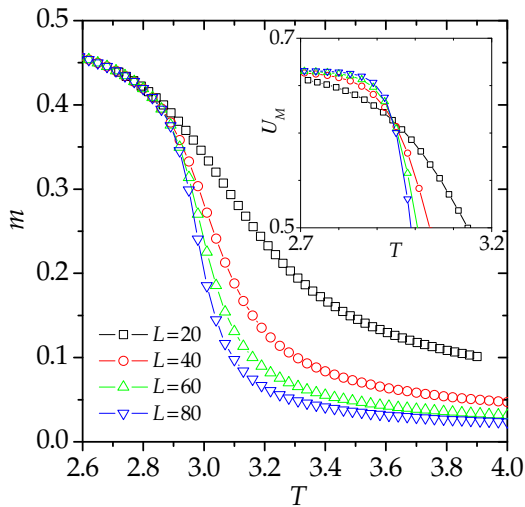}
\caption{Temperature dependence of the normalized magnetization and the Binder cumulant (inset),
for different lattice sizes as indicated.  Cumulants cross at around $T \approx 2.94$ and 
$U_M^* \approx 0.613$.}
\label{figure2} 
\end{center}
\end{figure}

As it will be justified below, the normalized magnetization $m$ is a good parameter to study
the magnetic order of the system.  Although the ground state of the present model is not purely ferromagnetic, it can be shown that $m$
tends to a constant value: $\lim_{T \to 0} m= 0.5$.
In Figure \ref{figure2} we show the temperature dependence of $m$ for different 
lattice sizes. In the inset, the corresponding Binder cumulants are presented, which intersect at around 
$T \approx 2.94$ and $U_M^* \approx 0.613$, confirming the existence of a finite-temperature 
continuous phase transition.

\begin{figure}[t]
\begin{center}
\includegraphics[width=7cm,clip=true]{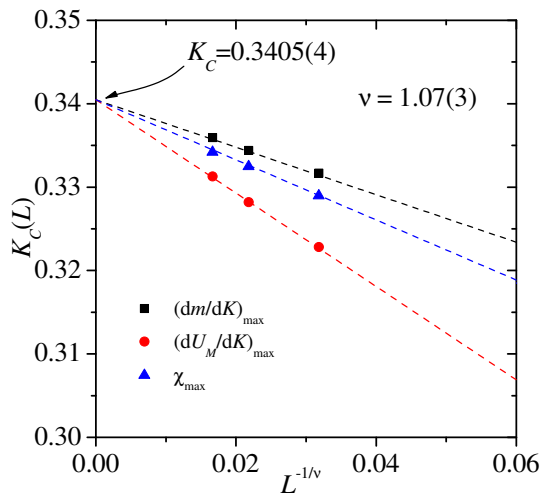}
\caption{Size-dependent inverse temperature, $K_c(L)$, to which the maxima of the derivatives 
of the order parameter, $\left(d m / d K \right)_{\rm max}$, and the Binder cumulant, 
$\left(d U_M / d K \right)_{\rm max}$, as well as of the susceptibility, $\chi_{\rm max}$, 
reach their maximum values. Dashed lines correspond to fitting results.}
\label{figure3} 
\end{center}
\end{figure}

Furthermore, finite-size scaling theory \cite{BinderBook,Fisher1971,Privman1990} allows for other efficient routes to estimate $T_c$ from the numerical data. One possible method, which is more accurate than the intersection of the Binder cumulant presented in the inset of Figure \ref{figure2}, relies on the extrapolation of the size-dependent inverse temperature, $K_c(L)$, to which different thermodynamic quantities reach their maximum values. Scaling theory predicts that 
\begin{equation}
K_c(L)=K_c+{\rm const}.L^{-1/\nu}. 
\label{KcL}
\end{equation}
Among others, the maxima of the slopes of the order parameter and the Binder cumulant,
$\left(d m / d K \right)_{\rm max}$ and $\left(d U_M / d K \right)_{\rm max}$, 
as well as of the susceptibility, $\chi_{\rm max}$, are quantities that can be used with this method.
Performing a simultaneous fitting procedure using Eq. (\ref{KcL}) 
and setting only two variables on the fit, i.e., $K_c$ and the exponent $\nu$,
we obtain $K_c=0.3405(4)$ or $T_c=2.937(3)$, and $\nu=1.07(3)$. Figure \ref{figure3} shows 
the corresponding plot of these quantities versus $L^{-1/\nu}$.  Note that the 
critical temperature coincides, within numerical errors, with the value calculated 
from the crossing of the cumulants.   

Next, to calculate precise values for the critical exponents (including $\nu$),
we make a conventional finite-size scaling analysis \cite{BinderBook,Fisher1971,Privman1990}.   
At criticality, the finite-size scaling relations are
\begin{equation}
C_H=L^{\alpha / \nu} \widetilde{C}_H(L^{1/\nu} \epsilon) \label{SRheat}
\end{equation}
\begin{equation}
m = L^{-\beta/\nu} \widetilde{m}(L^{1/\nu} \epsilon) \label{SRmagNor}
\end{equation}
\begin{equation}
\chi= L^{\gamma/\nu} \widetilde{\chi}(L^{1/\nu} \epsilon) \label{SRsusM}
\end{equation}
\begin{equation}
U_M= \widetilde{U}_M(L^{1/\nu} \epsilon) \label{SRcumM}
\end{equation}
for $L \to \infty$, $\epsilon \to 0$ such that $L^{1/\nu} \epsilon $ = finite, where 
$\epsilon \equiv T/T_c - 1$.  Here, $\alpha$, $\beta$, $\gamma$ and $\nu$ are the standard critical 
exponents of the specific heat ($C_H \sim |\epsilon|^{-\alpha}$ for $\epsilon \to 0$, $L \to \infty $), 
order parameter ($m \sim -\epsilon^{\beta} $ for $\epsilon \to 0^-$, $L \to \infty$),
susceptibility ($\chi \sim |\epsilon|^\gamma $ for $\epsilon \to 0$, $L \to \infty$) and
correlation length $\xi$ ($\xi \sim |\epsilon|^{-\nu}$ for $\epsilon \to 0, L \to \infty$), respectively.
$\widetilde{C}_H$, $\widetilde{m}$, $\widetilde{\chi}$ and $\widetilde{U}_M$ are scaling
functions for the respective quantities.

Following the line of Refs. \cite{Ferrenberg1991,Janke1994,Binder2001},
the critical exponent $\nu$ is firstly computed by considering different derivatives with respect to the inverse 
temperature $K=1/T$, for example, the derivative of the Binder cumulant and the logarithmic 
derivative of the order parameter.  It is expected that the maximum value of these derivatives 
as a function of the lattice size follows a power law of the form $\sim L^{1/\nu}$.  Once the value
of $\nu$ is known, the critical exponent $\gamma$ can be determined by scaling the maximum value 
of the susceptibility, i. e., from $\chi_{\mathrm{max}} \sim L^{\gamma/\nu}$.  In addition,
the standard way to extract the exponent $\beta$ is to study the scaling behavior of the 
order parameter at the point of inflection, $m_{\mathrm{inf}}$.  We expect that 
$m_{\mathrm{inf}} \sim L^{-\beta/\nu}$.  
From this analysis we obtain $\nu=1.02(2)$, $\gamma/\nu=1.746(6)$, and $\beta/\nu=0.126(6)$,
and then $\gamma=1.78(4)$  and $\beta=0.128(8)$. 
On the other hand, the maximum of the specific heat, 
$C_{H \mathrm{max}}$, does not follow a power law.  Instead, we observe a logarithmic 
divergence of the form $C_{H \mathrm{max}} \approx 0.071+0.268 \ln(L)$.  This implies that 
the corresponding critical exponent is zero, $\alpha=0$ \cite{StanleyBook}.  

\begin{figure}[t]
\begin{center}
\includegraphics[width=\linewidth,clip=true]{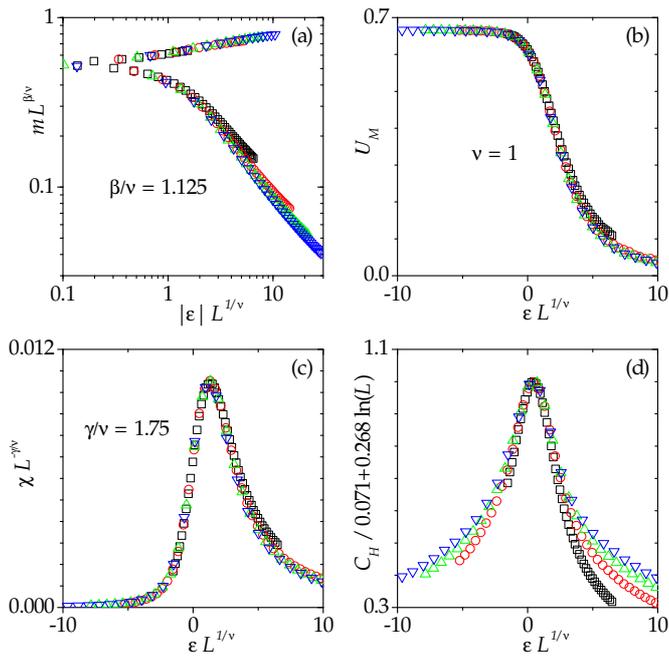}
\caption{Data collapses for (a) the order parameter, (b) the Binder cumulant, (c) the
susceptibility and (b) the heat capacity.  Symbols are the same 
as in Figure \ref{figure2}.}
\label{figure4} 
\end{center}
\end{figure}

We simulated system sizes only up to $L=80$, which gives probably the main contribution to the size of the error bars of the obtained critical exponents. Nevertheless, as the obtained values are compatible with those of the 2D Ising model, 
$\nu=1$, $\beta=1/8=0.125$, $\gamma=7/4=1.75$, and $\alpha=0$, we conclude that 
the observed phase transition probably belongs to this universality class.  We should notice that
$\nu=1.07(3)$, calculated from the simultaneous fitting to the Eq. (\ref{KcL}), 
is very close to one but does not agree within error bars. This discrepancy may be due to the 
finite-size effects.

Figures \ref{figure4} (a)-(d) show good 
data collapses for, respectively, the order parameter, the Binder cumulant, and the
susceptibility where we use the Ising exponents (the data collapses
do not change significantly if the exponents obtained numerically are used), 
while for the heat capacity we use the logarithmic correction term determined above.  
As we can see, we obtain very satisfactory scalings.

\subsection{Bond structure and the ground-state properties}

\begin{figure}[t]
\begin{center}
\includegraphics[width=7cm,clip=true]{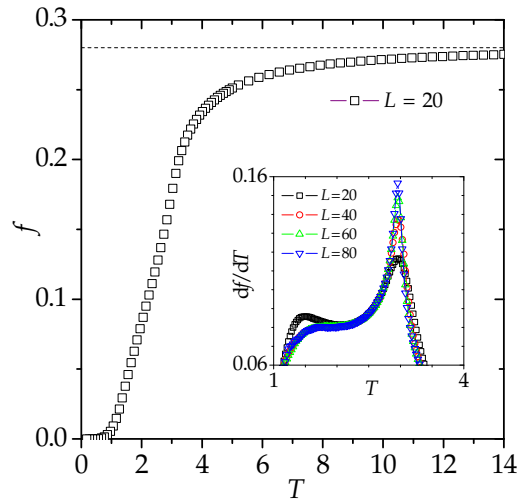}
\caption{Mean fraction of frustrated plaquettes as function of $T$ for a lattice size $L=20$.  
Inset: Temperature derivative $df/dT$ for different lattice sizes as indicated.  }
\label{figure5} 
\end{center}
\end{figure}

In the previous subsection we have shown that the present model has a 
standard finite-temperature transition. We show here that
the low-temperature equilibrium dynamics and the ground-state properties of the model present
a rather interesting and rich behavior. To carried out the corresponding studies, we have performed 
additional simulations on smaller lattice sizes (ranging from $L=6$ to $L=20$) and for lower 
temperature, $T_R=0.1$.  

Figure \ref{figure5} shows, for a system of lattice size $L=20$, 
the fraction of frustrated plaquettes, $f$, as function of $T$.  
For the limit $T \to \infty$ this fraction tends to a constant value of about $0.28$.   
In contrast, the Edwards-Anderson $\pm J$ model has a larger value $f=0.5$ (in this case,
for any temperature).  This shows that, even for the full-disordered state, the majority rule given by Eq. (\ref{rule})
induces strong correlations between bonds. On the other hand, for $T \to 0$ the model shows 
a non-frustrated ground state with $f=0$.  What happens is that, at equilibrium, the dynamics 
efficiently eliminates any frustration.   
  
In the inset of Figure \ref{figure5} we can see the temperature derivative $df/dT$ for different 
lattice sizes. Because it is not possible to equilibrate large lattice sizes up to 
very low temperatures,
these curves are only shown in the range of $T=1$ to $T=4$.  As expected, $df/dT$ shows a size-dependent 
peak at approximately $T_c$, but another at $T \approx 1.4$ which for $L > 20$ quickly leads 
to a characteristic knee feature.  We can not associate the latter temperature to a critical one.  Nevertheless, this is in the temperature range at which we observe the onset of the slow dynamics, as shown below.

\begin{figure}[t]
\begin{center}
\includegraphics[width=7cm,clip=true]{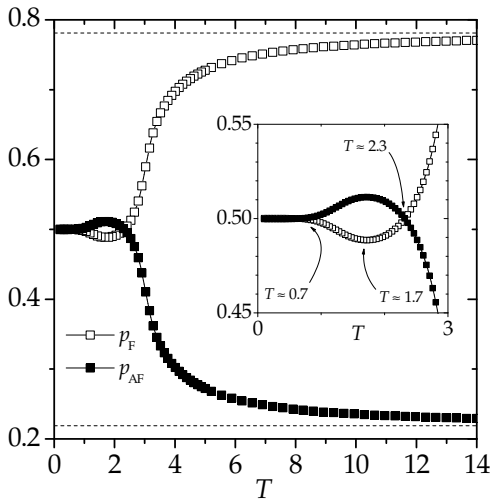}
\caption{The fractions of ferromagnetic and antiferromagnetic bonds $p_{\mathrm{F}}$ and 
$p_{\mathrm{AF}}$, respectively, as function of $T$. Dotted lines indicate the limit values of
these quantities when $T \to \infty$.  Inset shows the same curves at low temperatures.}
\label{figure6} 
\end{center}
\end{figure}

A more complex behavior is displayed by the fractions of ferromagnetic and 
antiferromagnetic bonds, $p_{\mathrm{F}}$ and $p_{\mathrm{AF}}$, respectively. Figure \ref{figure6} shows 
these quantities for $L=20$.  It is observed in Figure \ref{figure6} that when $T \to \infty$ the fractions of ferromagnetic and antiferromagnetic bonds reach a limiting constant value, with $p_{\mathrm{F}} > p_{\mathrm{AF}}$. 
Both limit values are easily calculated. The number of 
configurations of the spins surrounding a given pair is $2^6$.  Two of them corresponds to   
all spins pointing in the same direction while in another twelve, one spin point in a direction 
opposite to the remaining (e. g., one spin up and the other down, or vice versa).  The majority rule, Eq. (\ref{rule}),
indicates that for these fourteen configurations the bond $J_{ij}$ is antiferromagnetic and otherwise is ferromagnetic.  Then, 
considering that for $T \to \infty$ all configurations are equally likely, we obtain    
$\lim_{T \to \infty} p_{\mathrm{AF}} = 14/2^6=0.21875$ and 
$\lim_{T \to \infty} p_{\mathrm{F}}=1-0.21875=0.78125$. Dotted lines in Figure \ref{figure6} 
indicate these limit values.

Furthermore, it can be observed in Figure \ref{figure6} that for $T \to 0$ the system tends to a ground 
state with the same fractions of both types of bonds, 
$\lim_{T \to 0} p_{\mathrm{F}}=\lim_{T \to 0} p_{\mathrm{AF}}=0.5$. Interestingly enough, the
inset shows that the fractions of ferromagnetic and antiferromagnetic bonds have a non-monotonous behavior in a finite temperature range.   At $T \approx 2.3$ the curves of $p_{\mathrm{F}}$ and $p_{\mathrm{AF}}$ cross each other and, between this temperature and
$T \approx 0.7$, the fraction of ferromagnetic bonds is smaller than $0.5$ and reaches a minimum 
at $T \approx 1.7$ (and of course, $p_{\mathrm{AF}}$ reaches its maximum).  Then, for $T \lesssim 0.7$ 
both quantities are virtually identical.  In what follows, we will always have in mind these 
particular temperatures when analyzing other observables.   

\begin{figure}[t]
\begin{center}
\includegraphics[width=7cm,clip=true]{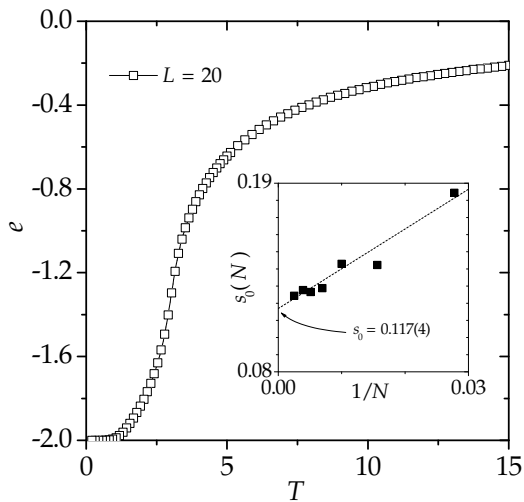}
\caption{Temperature dependence of the mean energy per spin for $L=20$.  Inset shows the 
ground-state entropy as function of $1/N$.}
\label{figure7} 
\end{center}
\end{figure}

Although the ground state has an equal number of ferromagnetic and antiferromagnetic bonds, 
there are important differences between this system and the Edwards-Anderson $\pm J$ model.
Figure \ref{figure7} shows the mean energy as a function of $T$.  We can see that this quantity
tends to $-2$ at zero temperature, $e_0 \equiv \lim_{T \to 0} e =-2$.  This is a logical limit value since,
as we have previously observed, the ground state is not frustrated.  In contrast, the 2D Edwards-Anderson $\pm J$ model
has a higher value of $e_0 \approx -1.4$ \cite{Roma2004}.    

We can obtain more information from the ground-state entropy per spin, $s_0$. In order to
determine $s_0$, we have simulated the system between $T_1=300$ and
$T_R=0.1$. Samples with $L \le 20$ were used because only for these lattice sizes it is possible to 
achieve equilibrium at such low temperatures.  The ground-state entropy per spin 
(in units of $k_B$) is calculated using the 
thermodynamic integration method \cite{Roma2004,Kirkpatrick1977,Binder1985} and is defined by 
the expression \cite{Perez2012}     
\begin{equation}
s_0(N)= \ln 2 + \int_{e_{\infty}}^{e_0} \frac{\mathrm{d}e}{T}  ,
\label{entro}
\end{equation}
where $e_{\infty} \equiv \lim_{T \to \infty} e = 0$. Inset in Figure \ref{figure7} shows $s_0(N)$ 
as function of $1/N$.  Extrapolating, we obtain a thermodynamic limit value of 
$s_0 \equiv \lim_{N \to \infty} s_0(N)=0.117(4)$.  To extrapolate, only lattices with $L \ge 10$ were
used because the entropy for smaller lattice sizes present some oscillations, which seem to be related to the
periodic boundary conditions \cite{Perez2012}.  This entropy value is larger than the corresponding 
one of the 2D Edwards-Anderson $\pm J$, $s_0^{\mathrm{EA}} \approx 0.07$ 
\cite{Perez2012,Saul1994,Zhan2000,Hartmann2001b}.  So that, our system has a non-frustrated and 
highly-degenerated ground state.  
 
\begin{figure}[t!]
\begin{center}
\includegraphics[width=\linewidth,clip=true]{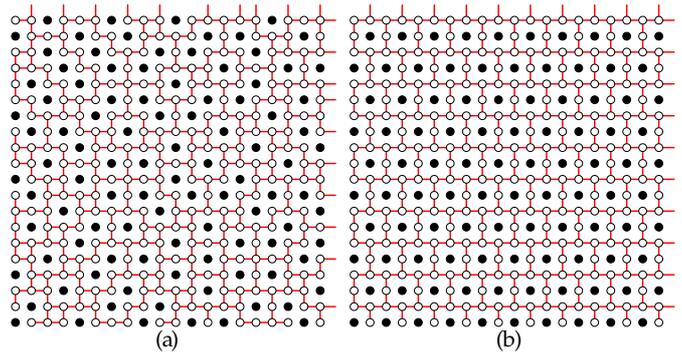}
\caption{(a) Typical and (b) fully ordered ground-state configurations for lattice size $L=20$.  
Minority and majority spins are indicated by closed and open black circles, respectively, 
while ferromagnetic bonds are denoted by red lines (antiferromagnetic bonds are omitted for simplicity).}
\label{figure8} 
\end{center}
\end{figure}

A better understanding of this phenomenon can be achieved by analyzing the structure of the 
ground state.  The parallel-tempering algorithm can be used as a heuristic to obtain
the lowest energy configurations \cite{Moreno2003,Roma2009}.  We emphasize that for this application it is not 
necessary to reach equilibrium, because only configurations with $e=-2$ are sought.
Figure \ref{figure8} (a) shows a typical ground state.  Analyzing many configurations 
like the one in Figure \ref{figure8} (a), one can conceive that a fully ordered structure as the one shown in Figure \ref{figure8} (b) might also be found in the ground-state. Although this latter configuration has energy $e=-2$ and is compatible with the majority rule, Eq. (\ref{rule}), it is extremely rare and is very unlikely to obtain using any algorithm.

\begin{figure}[t]
\begin{center}
\includegraphics[width=\linewidth,clip=true]{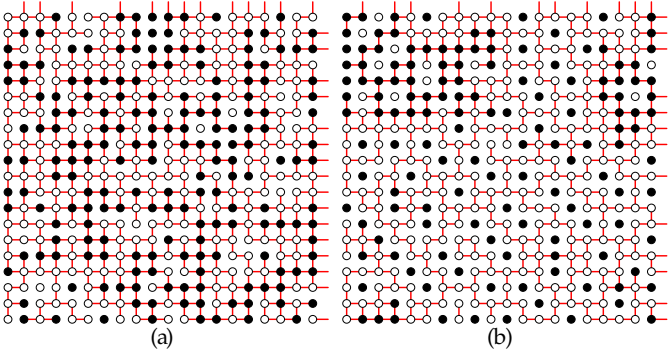}
\caption{Typical configurations for (a) $T=\infty$ and (b) $T=3.0$. Symbols are the same as 
in Figure \ref{figure8}.}
\label{figure9}
\end{center}
\end{figure}

Such ground-state configurations have some interesting features.  On one hand, we can see that
a quarter of the spins (minority spins) point in a direction opposite to the remaining spins (majority spins).
This is clearly related to the fact that $\lim_{T \to 0} m= 0.5$ and the reason for using the magnetization $m$ as a good order parameter. More strictly, the ground state is not fully ferromagnetic but has a ferrimagnetic character. On the other hand, notice that 
minority spins have as nearest neighbors majority spins only, and these pairs are linked by 
antiferromagnetic bonds. Thus, in Figs. \ref{figure8} (a) and (b) minority spins look as if they were isolated, i.e. minority spins do not interact among them.

\begin{figure}[t!]
\begin{center}
\includegraphics[width=7cm,clip=true]{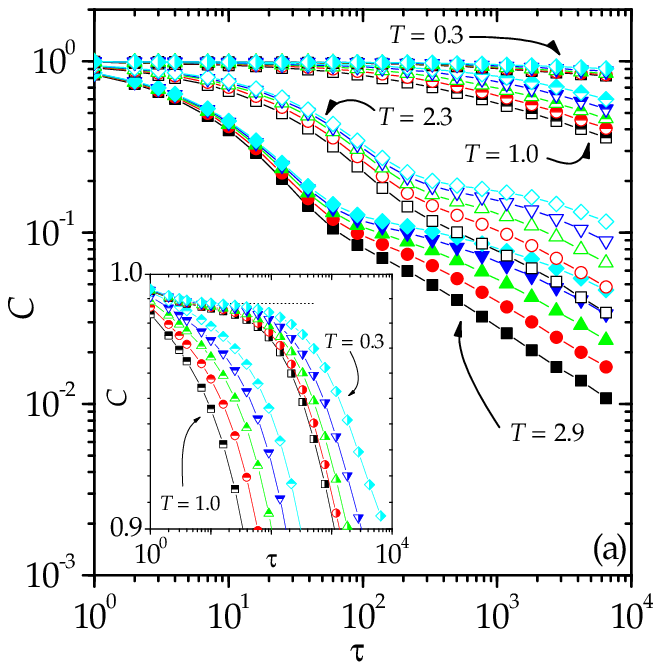}
\includegraphics[width=7cm,clip=true]{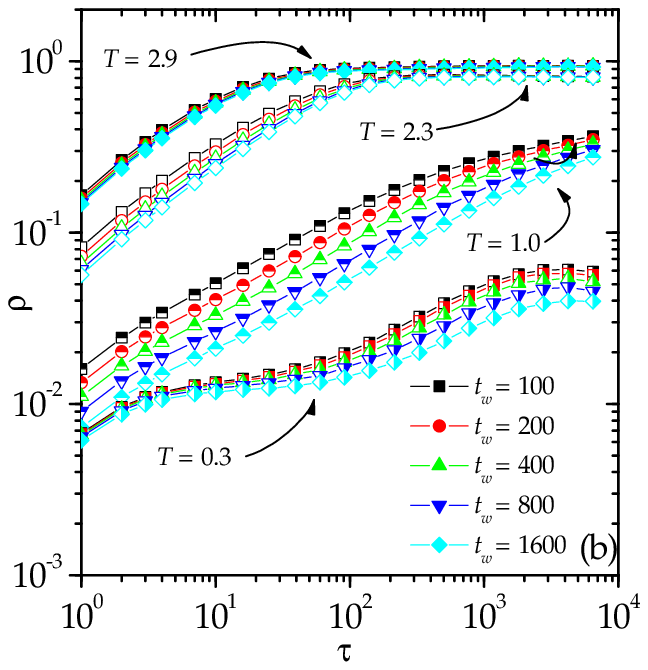}
\caption{(a) Spin correlation and (b) spin integrated response functions, at temperatures 
$T=2.9$, $2.3$, $1.0$, and $0.3$, and for different $t_w$ as indicated.  Inset shows a zoom 
of the correlations functions for the two lower temperatures.}
\label{figure10}
\end{center}
\end{figure}

At infinite temperature all configurations are equally likely.  Figure \ref{figure9} (a)
shows a typical high-temperature configuration, which has equal number of up and down spins (here, closed 
and open black circles represent, respectively, up and down spins).  But, for a 
temperature close to the critical one, $T=3.0$, Figure \ref{figure9} (b) shows that
the ``low-temperature phase'' begins to develop [compare with Figures \ref{figure8} (a) and (b)].  We can 
see that excitations are characterized by violations of the above features observed in the 
ground state (let us recall that, at $T=0$, minority spins have as nearest neighbors majority spins only, 
and these pairs are linked by antiferromagnetic bonds).  For the particular temperatures 
indicated in the inset of Figure \ref{figure6}, in the range between $T = 2.3$ and $T = 0.7$, 
no anomaly is observed and all configurations are similar to those of the ground state.  

\begin{figure}[t]
\begin{center}
\includegraphics[width=7cm,clip=true]{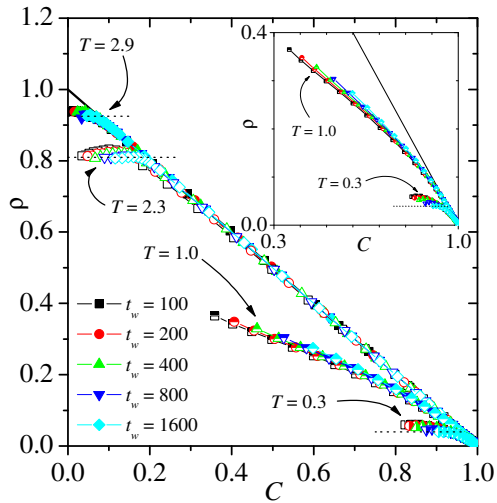}
\caption{FDT parametric plots $\rho(c)$ for the cases shown in Figs. \ref{figure10} (a) and (b).
Inset shows a zoom of the FDT plots for the two lower temperatures.}
\label{figure11}
\end{center}
\end{figure}

\subsection{Out-of-equilibrium dynamic}

At very low temperatures, it is numerically hard to reach equilibrium for large lattice sizes.  
In order to explore this region, we carry out
extensive out-of-equilibrium simulations.  Lattices of $L=200$ and six values of the waiting time, 
$t_w$ = 100, 200, 400, 800, and 1600, were used. For each temperature we have performed 
averages over $5000$ independent runs. 

Figures \ref{figure10} (a) and (b) show the spin correlation and spin integrated response 
functions for different temperatures. For temperatures $T=2.9$ (just below $T_c$) and $T=2.3$, 
above which $p_{\mathrm{F}} > p_{\mathrm{AF}}$, correlation functions tend to develop a plateau 
for increasing waiting time while response functions get stuck in a well defined plateau for all waiting times. 
When represented in a FDT parametric plot, as in Figure \ref{figure11}, this behavior leads to the 
typical coarsening-like violation of the FDT \cite{Crisanti2003} signaled by a constant $\rho(C)$ curve, 
highlighted with dotted lines in Figure \ref{figure11}. This would be the expected result for our model, 
which undergoes a thermodynamic continuous phase transition belonging to the 2D Ising universality class.

For a lower temperature $T=1.0$, all correlation and response functions curves in Figure \ref{figure10} 
show a slightly different behavior which is more evident in Figure \ref{figure11}. In this case the 
FDT parametric plot seems to slowly converge to a FDT violation characterized by a finite slope 
at sufficiently long times, i.e. small values of $C$, although some curvature due to finite $t_w$ effects cannot be discarded.  Interestingly, this FDT violation is reminiscent of the typical FDT violation 
found in mean-field models for structural glasses \cite{Crisanti2003,Berthier2011}, characterized 
by a one-step replica symmetry breaking \cite{Parisi2003}.
This singular behavior is displayed within a wide temperature range around $T=1.0$, 
and is concomitant with the onset of the slow dynamics.  On the other hand, even more surprising is 
the fact that upon further lowering the temperature, for $T=0.3$ the correlation functions show the 
emergence of a plateau close to $C=1$ and the response function also develops a plateau at a very small value, 
as shown in Figure \ref{figure10}. The result is a coarsening-like FDT violation as shown in Figure \ref{figure11}.

\begin{figure}[t]
\begin{center}
\includegraphics[width=7cm,clip=true]{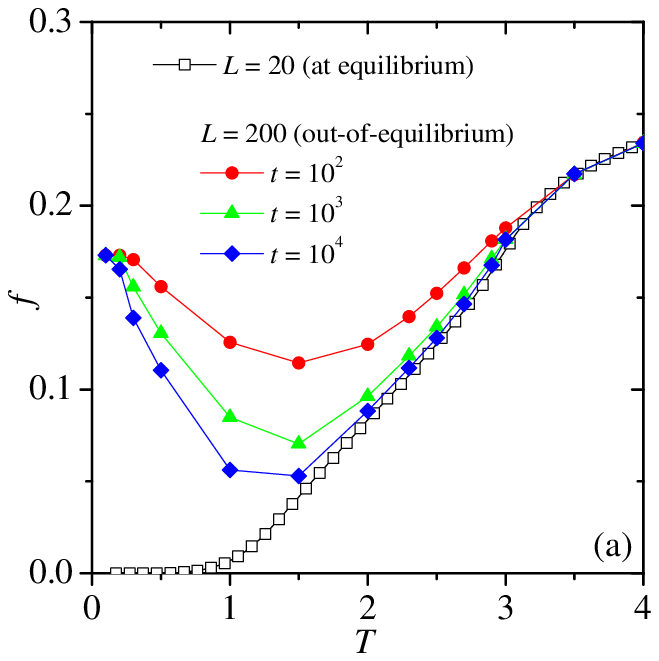}
\includegraphics[width=7cm,clip=true]{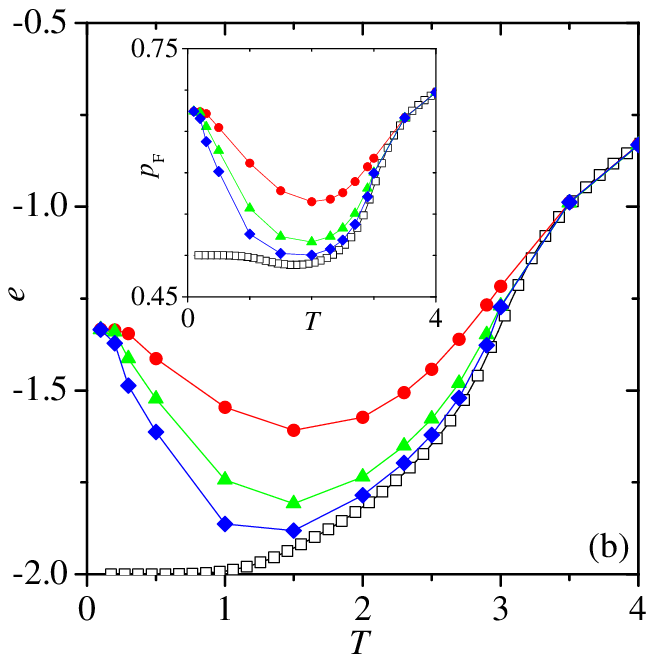}
\caption{Temperature dependence of (a) the mean fraction of frustrated plaquettes 
and (b) the mean energy per spin.  Both panels show the corresponding equilibrium 
and out-of-equilibrium curves for different times as indicated. Inset shows the same
information for the mean fraction of ferromagnetic bonds.}
\label{figure12}
\end{center}
\end{figure}

In order to shed some light on this phenomenon, we analyze again the bond structure 
but this time comparing thermal equilibrium with out-of-equilibrium measurements 
of the same observables. Figure \ref{figure12} (a) shows the temperature dependence 
of the mean fraction of frustrated plaquettes comparing equilibrium with time-dependent 
measurements. The same comparison is shown in Figure \ref{figure12} (b) and its inset for 
the mean energy per spin and the mean fraction of ferromagnetic bonds, respectively. 
From these figures some common features are revealed. For temperatures around the critical 
temperature $T_c$, all quantities rapidly relax to their equilibrium values [for long 
times there is not a perfect agreement between equilibrium and out-of-equilibrium values, 
because both calculations were performed for different lattice sizes, as indicated in the key 
of Figure \ref{figure12} (a)]. Therefore, in this temperature range, for example at $T=2.9$ and $T=2.3$, 
the out-of-equilibrium dynamics is fast enough to develop a coarsening process, as noted in 
the analysis of the QFDT, Figure \ref{figure11} above.

A different situation arises for $T \sim 1$, where for the studied time scales the 
dynamics is clearly slower and the system remains in a state with higher energy than 
expected at equilibrium. In this state, as can be observed in Figure \ref{figure12} (a), 
the slow dynamics has not been effective in removing the frustration of the system. 
It is not surprising then that this dynamic and structural changes around $T \sim 1$ 
lead to a different QFDT characteristic, which in this case resembles the one 
observed in structural glasses.

Finally, at very small temperatures, Figs. \ref{figure12} (a) and \ref{figure12} (b) show
that the very slow dynamics is not able to remove the frustration and the system is 
stuck in a high energy configuration. However, the values of mean fraction of frustrated 
plaquettes and mean energy per spin, let say for $T=0.3$, are very similar to those measured 
for temperatures close to $T_c$, let say $T=2.9$. Our interpretation is that for these 
values of frustration and energy the relaxation processes should be very similar and then 
we observe the characteristic coarsening feature in the QFDT at $T=0.3$, Figure \ref{figure11}. 
At such a small temperature the relaxation is so slow that a different relaxation mechanism 
at longer times can not be discarded only based on our numerical simulations.

\section{Conclusions \label{sec:conc}} 

In this work we have studied the critical behavior, the ground-state properties 
and the out-of-equilibrium dynamics of a 2D Ising model with non-static interactions.
Its most important feature is that bonds are drawn according to a majority rule,
Eq. (\ref{rule}), which was designed to prevent the system from ordering 
in full ferromagnetic or antiferromagnetic states. Nevertheless, 
we have demonstrated through equilibrium simulations, that the model 
has a low-temperature ferrimagnetic state and undergoes a continuous phase transition 
at $T_c=2.937(3)$, which probably belongs to the 2D Ising universality class.        

The ground state is nontrivial.  Although frustration is completely removed at $T=0$,
(i. e. the energy per spin is $e_0 =-2$) the ground state is highly degenerated 
with an entropy per spin of $s_0=0.117(4)$.  Interactions form a degenerate structure where
minority spins have as nearest neighbors majority spins only, and these pairs are 
linked by antiferromagnetic bonds.

This degenerate low-temperature phase has a low free energy  
and therefore is more stable than the one of the Ising model (which has a non-degenerate ground state).     
As a consequence, the critical temperature of the system, $T_c=2.937(3)$, is slightly larger 
than the one of the Ising ferromagnetic model, $T_c^\mathrm{Ising} \approx 2.269$ \cite{Onsager1944}. 

In order to quantify this statement we use a free-energy minimization 
criterion \cite{Roma2003}. Let us consider a generic model undergoing a continuous phase transition.
In the limits $T \to \infty$ and $T \to 0$ such a system is characterized by two extreme states: 
a fully-disordered (FD) state characterized, respectively, with an energy and an entropy 
per spin $e'_\mathrm{FD}$ and $s'_\mathrm{FD}$, and 
a fully-ordered (FO) state with the corresponding parameters $e'_\mathrm{FO}$ and $s'_\mathrm{FO}$.   
The Helmholtz free energy of these states are $f'_\mathrm{FD}=e'_\mathrm{FD} - T s'_\mathrm{FD}$ 
and $f'_\mathrm{FO}=e'_\mathrm{FO} - T s'_\mathrm{FO}$.  
The minimization criterion consists in considering that the critical behavior 
is mainly determined by the FD and FO states.  Since in thermodynamic equilibrium
the stable state corresponds to the minimum of the Helmholtz free energy, then
\begin{eqnarray}      
f'_\mathrm{FD} &<& f'_\mathrm{FO} \ \ \mathrm{if} \ T > T'_c \\
f'_\mathrm{FD} &=& f'_\mathrm{FO} \ \ \mathrm{if} \ T = T'_c \label{crit} \\
f'_\mathrm{FD} &>& f'_\mathrm{FO} \ \ \mathrm{if} \ T < T'_c .
\end{eqnarray}
Finally, from Eq. (\ref{crit}) we can estimate the critical temperature as
\begin{equation}
T'_c = \frac{\Delta e'}{\Delta s'}, \label{Tc2}
\end{equation}
where $\Delta e'=e'_\mathrm{FD} - e'_\mathrm{FO}$ and $\Delta s'=s'_\mathrm{FD} - s'_\mathrm{FO}$.

Our model and the Ising model have non-frustrated ground states and therefore 
$\Delta e=\Delta e^\mathrm{Ising}=2$.
However, since for the FD state both systems have the same entropy, 
$s_\mathrm{FD}=s^\mathrm{Ising}_\mathrm{FD}=\ln 2$, but for the FO state 
$s_\mathrm{FO}>s^\mathrm{Ising}_\mathrm{FO}$=0, then their entropy changes are 
different, $\Delta s<\Delta s^\mathrm{Ising}=\ln 2$. 
Hence, from Eq. (\ref{Tc2}) we can determine that the critical temperatures 
will be $T_c>T_c^\mathrm{Ising}$. 
Note that the free-energy minimization criterion does not provide a good estimation of
$T_c$, but it offers a very useful tool for understanding the critical behavior of a system 
with respect to parameter variations \cite{Roma2003}.

Finally, we have determined that the out-of-equilibrium dynamics has a novel behavior:
at very low temperatures as well as near $T_c$ (but below of this critical temperature)
we observed a coarsening-like FDT violation, while at an intermediate temperatures ($T \sim 1$)
the FDT parametric plots are quite similar to those found for structural glasses. 
Nevertheless, for the latter case, note that in the parametric FDT plot
the slope at sufficiently long times (FDT ratio) shows a small dependence on $t_w$, and therefore 
we cannot exclude that such phenomenon is due to finite-time effects.
Comparing equilibrium with out-of-equilibrium measurements of different observables 
($f$, $e$, and $p_{\mathrm{F}}$), we have been able to establish a connection between
this phenomenon and the bond structure properties of the system.  

O.A.P. acknowledges financial support from Universidad Nacional de Santiago del Estero, 
Argentina, under project CICyT-UNSE 23 A173. 
F.R. acknowledges financial support from CONICET (PIP 114-201001-00172) and Universidad
Nacional de San Luis, Argentina (PROIPRO 3-1-0214) and thanks the LPTHE, France, for 
hospitality during the preparation of this work. S.B. is partially supported by CONICET
(PIP11220090100051) and FONCyT (PICT2010-889).


\end{document}